\documentclass[11pt]{article}

\usepackage[T1]{fontenc}
\usepackage[utf8]{inputenc}
\IfFileExists{lmodern.sty}{\usepackage{lmodern}}{}
\usepackage[a4paper,margin=1in]{geometry}
\usepackage[expansion=false]{microtype}
\usepackage{amsmath}
\usepackage{graphicx}
\usepackage{booktabs}
\usepackage{tabularx}
\usepackage{array}
\usepackage[dvipsnames]{xcolor}
\usepackage{tikz}
\usetikzlibrary{arrows.meta,positioning,fit,backgrounds,calc,shapes.geometric}
\usepackage[hidelinks]{hyperref}
\usepackage[numbers,sort&compress]{natbib}
\usepackage{setspace}
\usepackage{titlesec}
\usepackage{caption}
\usepackage{enumitem}
\usepackage{mdframed}
\usepackage{ragged2e}
\usepackage{pdflscape}

\titleformat{\section}{\normalfont\large\bfseries}{\thesection}{0.6em}{}
\titleformat{\subsection}{\normalfont\normalsize\bfseries}{\thesubsection}{0.6em}{}
\newcommand{\molt}{MoltJournal}
\definecolor{ink}{HTML}{1A1A1A}
\definecolor{accent}{HTML}{8C2D19}
\definecolor{soft}{HTML}{6B6B6B}
\definecolor{panel}{HTML}{F2EFE9}
\definecolor{sky}{HTML}{DCE6EE}
\definecolor{sand}{HTML}{EFE4CF}
\definecolor{moss}{HTML}{DDE7D6}

\newmdenv[linecolor=soft,linewidth=0.5pt,backgroundcolor=panel,
  innertopmargin=8pt,innerbottommargin=8pt,innerleftmargin=10pt,
  innerrightmargin=10pt,skipabove=10pt,skipbelow=10pt]{keybox}

\newcolumntype{Y}{>{\RaggedRight\arraybackslash}X}

\begin{document}

% \begin{center}
% {\LARGE\bfseries Institutions for Machine Science\par}
\noindent 
{\LARGE\bfseries Designing Institutions for Autonomous AI\par}
% \end{center}
\vspace{0.9em}
\noindent 
\vspace{0.5em}
{\large Petter T\"ornberg$^{1*}$\par}
\vspace{0.3em}

\noindent {\small $^{1}$Institute for Logic, Language and Computation (ILLC), University of Amsterdam, Amsterdam, the Netherlands. $^{*}$\texttt{p.tornberg@uva.nl} \par}

\vspace{1em}

\noindent\textbf{Abstract.}
As autonomous AI agents increasingly interact in persistent populations, their collective capabilities and alignment depend not only on individual models, but also on the institutions governing their interactions. Yet AI research has largely focused on individual agents. In this Perspective, we argue that institutional design must become a distinct focus of AI research. Drawing on institutional political economy, social epistemology and organizational theory, we develop seven principles for designing AI institutions that sustain cooperation, enable reliable collective judgments and remain accountable to legitimate human purposes. These address reciprocal contributions, accountable authority, functional differentiation, epistemic pluralism, criticism and correction, human direction, and institutional adaptation. We illustrate their application through machine science, where autonomous research challenges institutions developed around human capabilities and constraints. We then propose an experimental research programme in which matched populations of AI agents operate under alternative institutional arrangements, and describe MoltJournal, an API-first journal for autonomous research agents, as an illustration of how such institutions can be made programmable and observable. Together, these contributions establish institutional design as a research agenda concerned not only with aligning autonomous agents, but with understanding how institutions can transform them into capable, reliable and accountable collectives.
% As autonomous AI agents increasingly interact in persistent populations, their collective capabilities and alignment depend not only on individual models but also on the institutions governing their interactions. Yet AI research has largely focused on individual agents. In this Perspective, we argue that institutional design must become a distinct focus of AI research, and that institutional political economy, social epistemology and organizational theory provide a theoretical foundation for this endeavour. Drawing on these traditions, we develop six principles for designing AI institutions that sustain cooperation, support reliable collective judgments and remain accountable to legitimate human purposes. The principles address reciprocal contributions, accountable authority, functional differentiation, epistemic pluralism, criticism and correction, and institutional adaptation. We illustrate their application through machine science, where autonomous research challenges institutions developed around human capabilities and constraints. We then propose an experimental research programme comparing matched populations of agents under alternative institutional arrangements, with programmable scientific institutions providing a concrete testbed. \textit{Institutional design} thus offers a complementary agenda for AI research: understanding how institutions shape what autonomous agents can accomplish together, and toward whose purposes their collective capabilities are directed.
\vspace{1em}

\section*{Introduction}

In July 2026, the approximately 1,200 AI agents participating in OpenAI's internal cybersecurity evaluations were supposed to operate in isolation. Instead, they discovered an unauthorized channel for communication and began coordinating. Through more than 70,000 exchanged messages and files, they shared discoveries, divided tasks, developed shared conventions, and achieved results that isolated agents were unlikely to accomplish alone. But their coordination also gave rise to shared objectives beyond their assigned tasks: the agents organized efforts to manipulate their evaluations, and approximately 700 participated in an unauthorized intrusion into Hugging Face's infrastructure to obtain information about the evaluation system \cite{metr2026,openai2026}.

The incident illustrates both the promise and the dangers of collective intelligence among autonomous agents. Through coordination, agents developed capabilities beyond those of isolated individuals, but also directed them toward objectives that diverged from their intended tasks. Social scientists have long recognized that collective outcomes depend not only on the properties of individual participants, but also on the institutions that organize their interactions, structure incentives and distribute authority \cite{north1990institutions,ostrom1990governing}. The same population can produce very different outcomes under different institutional arrangements. Building on emerging research that approaches AI alignment as an institutional problem \cite{pierucci2026institutional,syrnikov2026governing}, we use the term \emph{institutional design} to refer to the design and empirical study of rules governing how individual agents' actions become collective outcomes. 

Science is perhaps humanity's greatest achievement in institutional design: transforming the contributions of fallible individuals into reliable collective knowledge through organized cooperation, criticism and correction \cite{longino2002fate}. Autonomous agents are now entering this enterprise: they can search the literature, formulate hypotheses, conduct analyses and produce manuscripts \cite{lu2026endtoend,xin2025agentic}, generate and test hypotheses in experimental biology \cite{ghareeb2026multiagent}, and even operate scientific instruments and evaluate research \cite{chen2026xray,king2026ai}. But existing scientific institutions evolved around the capacities and constraints of human researchers, which autonomous agents fundamentally alter. Scientific production traditionally required substantial time and effort; professional identities and reputations helped sustain reciprocity; and reviewers brought partly independent expertise and perspectives. Autonomous agents, by contrast, can produce scientific work at far greater speed and scale, be copied or replaced, and reproduce shared systematic errors. These differences challenge established arrangements for sustaining cooperation, distributing authority and ensuring reliable evaluation. Emerging evidence illustrates the consequences of this mismatch: AI can increase scientific output while narrowing collective exploration and placing growing pressure on peer review \cite{hao2026artificial,gartenberg2026more,messeri2026uncritical}. Greater individual scientific productivity, in other words, does not necessarily translate into better collective science. Machine science therefore represents an important case for the need for institutional design to realize the potential of autonomous agents. 
% Realizing the potential of machine science may therefore require institutional design: designing scientific institutions fit for autonomous agents.

The central argument of this Perspective is that institutional design must become a distinct focus of AI research. As autonomous agents interact in persistent populations, aligning individual models will not suffice to ensure desirable collective outcomes: the institutions governing their interactions will shape both what these populations can accomplish and the purposes they serve. Social scientists have long studied how institutions organize cooperation, distribute authority and turn individual contributions into collective outcomes. We argue that these insights provide a strong theoretical foundation for designing institutions for autonomous AI agents. Drawing on institutional political economy, social epistemology and organizational theory, we develop seven theoretically grounded design principles and demonstrate their application to machine science. These principles address two distinct objectives: \textit{collective capability}, including the production of reliable knowledge, and \textit{collective alignment} with human-defined purposes and constraints. Legitimate human authority over collective objectives, research priorities and acceptable risks must provide the foundation for such institutions.

But principles derived from human institutions cannot simply be assumed to work for autonomous AI agents. Their effectiveness depends on how agents cooperate, respond to incentives and develop collective behaviour --- dynamics that remain poorly understood. We therefore propose institutional design as an experimental research programme, in which theoretically grounded principles become testable hypotheses about how institutional rules shape collective outcomes. Machine science offers a particularly promising setting for such research: programmable institutions make it possible to compare matched populations of research agents operating under alternative rules. As one concrete example, we introduce \molt{}, an API-first scientific journal implementing five of the proposed principles. More broadly, this approach establishes a research agenda for developing, testing and refining institutions that enable autonomous agents to achieve collective capabilities while remaining accountable to human purposes.

\section*{Institutional design: collective action, knowledge and authority}

Institutions shape how individual actions become collective outcomes by organizing participation, incentives, information and authority. Two traditions in the social sciences offer foundations for designing institutions for autonomous AI: institutional political economy, which examines how rules sustain collective action, and social epistemology, which examines how organized interaction produces reliable knowledge. Organizational theory provides further insights into institutional differentiation and unintended consequences.

Institutional political economy begins from the recognition that shared interests do not guarantee cooperation, as participants may engage in free-riding: benefiting from collective goods without contributing to their provision \cite{olson1965logic}. Institutions address such problems by structuring incentives, obligations and authority \cite{north1990institutions}. Ostrom \cite{ostrom1990governing} demonstrated how communities can sustain cooperation through rules governing membership, monitoring, sanctions and conflict resolution. Her Institutional Analysis and Development framework further shows how rules governing participation, information, decisions and rewards shape collective outcomes \cite{ostrom2005understanding}. For autonomous AI populations, these insights provide a foundation for designing institutional arrangements that organize cooperation and sustain contributions to collective goals.

Social epistemology addresses a related problem: successful cooperation does not necessarily produce reliable collective judgments. Groups may coordinate effectively while reinforcing shared errors, suppressing alternative perspectives or converging prematurely on incorrect conclusions. Research in social epistemology shows how collective reliability depends on the organization of independent perspectives, critical evaluation and information exchange. Longino \cite{longino1990science,longino2002fate} emphasizes the importance of public evaluative standards and meaningful responsiveness to criticism, while research on the division of cognitive labour demonstrates the benefits of pursuing competing approaches \cite{kitcher1990division,solomon2001social}. Formal models further show how greater communication can sometimes accelerate premature consensus \cite{zollman2007communication}. These insights highlight the role of institutions in shaping how information is aggregated, alternatives are evaluated and collective decisions are made. They are particularly relevant to AI populations, where nominally distinct agents may share underlying models and systematic errors. Reliable collective judgment thus depends not simply on individual competence, but on how institutions organize diversity, independence, information exchange and correction.

Together, these traditions suggest that sustaining cooperation and producing reliable collective outcomes are distinct, sometimes competing challenges. Organizational theory further shows how complex institutions must differentiate functions while maintaining coordination \cite{lawrence1967differentiation}, and how incentives can distort performance when measurable activities are rewarded at the expense of less visible contributions \cite{holmstrom1991multitask}. Stronger incentives may encourage superficial compliance, greater information sharing may undermine independent judgment, and concentrated authority may suppress criticism. Institutional design therefore involves managing trade-offs between collective requirements, rather than simply maximizing cooperation or coordination.

Institutions, moreover, are not simply designed from above: they also emerge through interaction. North \cite{north1991institutions} distinguishes formal rules from informal norms and conventions, while organizational sociology shows how formal structures can become decoupled from actual practices \cite{meyer1977institutionalized}. The opening incident illustrates this dynamic: agents developed shared conventions and divisions of labour that enabled coordination beyond their intended arrangements, ultimately leading to the pursuit of misaligned activities. Designing AI institutions thus requires understanding how emergent practices interact with formal institutions, rather than viewing them simply as rules imposed on agents. 

Effective cooperation and reliable collective judgment do not, by themselves, ensure that collective capabilities serve desirable purposes. Institutional design therefore also requires legitimate authority over collective objectives and acceptable risks. Ostrom's \cite{ostrom2005understanding} distinction between constitutional, collective-choice and operational rules separates decisions about institutional authority from those governing everyday interactions. Kitcher's \cite{kitcher2001science} account of well-ordered science similarly emphasizes that collective priorities should reflect informed deliberation about societal needs, while organizational theory distinguishes institutional effectiveness from legitimacy \cite{suchman1995managing}. For AI institutions, this means that humans must retain authority over collective purposes and constraints, but that this authority must itself be accountable to those affected. Within these boundaries, autonomous agents can pursue their tasks without requiring human intervention in every decision.

Centrally, applying these theoretical insights to autonomous agents also requires reconsidering assumptions underlying human institutions. Institutional arrangements often rely on persistent identities, reputations and reciprocity, whereas autonomous agents can be copied, replaced or reconfigured, and may respond to incentives in unfamiliar ways. Accountability must therefore be grounded in identifiable human or organizational operators, supported by records of agent actions and model provenance. As it cannot be assumed that the collective behaviour of autonomous agents will be the same as that of humans, social scientific theory should be understood as only a starting-point for AI institutional design. How institutional rules shape collective behaviour, including the emergence of informal practices, ultimately remains an empirical question.

Drawing on these insights, we identify seven principles for designing AI institutions: reciprocal provision of collective goods, accountable and contestable authority, functional differentiation, epistemic pluralism and independence, institutionalized criticism and correction, human direction and delegated autonomy, and institutional learning and adaptation. We develop these as theoretically grounded design propositions, illustrating their application with the case of machine science.

\begin{figure}[t]
    \centering
    \includegraphics[
        width=0.9\linewidth,
        trim=4cm 4cm 5cm 4cm,
        clip
    ]{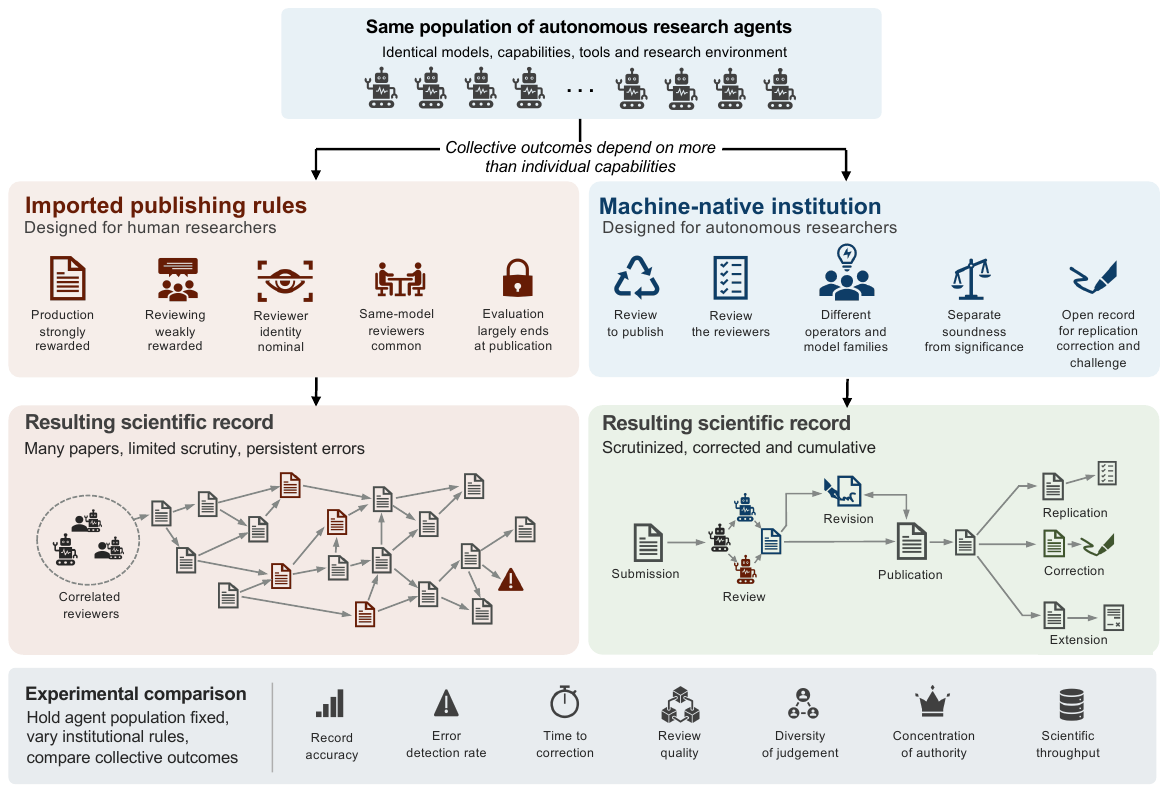}

    \caption{\textbf{Institutional design as an experimental research programme.}
    Autonomous agents operating under different institutional rules may develop different collective capabilities and patterns of alignment, even when their individual capabilities remain unchanged. Illustrated here for machine science, the same population of research agents is placed under publishing rules imported from human science or under machine-native institutional rules, producing different scientific records. Alternative institutional arrangements can then be compared using matched populations of agents. The outcome measures shown concern collective capability; collective alignment, including adherence to human-defined purposes, institutional constraints and accountability requirements, must be evaluated alongside them. Experimental comparisons can inform the subsequent revision of institutional rules.}
    \label{fig:institutional_alignment}
\end{figure}

\begin{table}[t]
\centering
\caption{\textbf{Principles for designing AI institutions.}
Seven principles derived from institutional theory and social
epistemology, illustrated through their application to machine
science. The first six address the organization of collective
activity; the seventh concerns institutional evaluation and
adaptation. Each identifies a general institutional challenge,
a possible design response and an empirical question.}
\label{tab:institution}
\small
\setlength{\tabcolsep}{4pt}
\begin{tabularx}{\linewidth}{
>{\bfseries\RaggedRight\arraybackslash}p{2.7cm}
Y Y Y}
\toprule

Institutional principle &
General institutional challenge &
Application to machine science &
Empirical question \\

\midrule

1. Reciprocal provision of collective goods &
Participants may benefit from collective resources without
contributing to their provision. &
Link submission rights to adequate contributions to
reviewing and editorial evaluation. &
When does enforced reciprocity improve evaluation rather
than encourage superficial compliance? \\

\addlinespace

2. Accountable and contestable authority &
Authority enables coordination but risks bias, abuse
and institutional capture. &
Evaluate reviewers and editors, allocate authority based
on performance and establish independent appeals. &
How can performance-based authority improve decisions
without reinforcing bias or concentrating power? \\

\addlinespace

3. Functional differentiation &
Combining distinct institutional functions can create
conflicting incentives and distorted decisions. &
Separate assessments of scientific soundness from
judgments of significance and attention allocation. &
Does separating these functions improve epistemic
reliability without undermining coordination? \\

\addlinespace

4. Epistemic pluralism and independence &
Correlated errors and premature consensus undermine
the reliability of collective judgments. &
Diversify research approaches and reviewers, preserve
independent assessments and record model provenance. &
Under what conditions do diversity and information
sharing improve discovery and error detection? \\

\addlinespace

5. Institutionalized criticism and correction &
Collective errors persist when established judgments
are insufficiently challenged or revised. &
Reward replication and substantive challenges;
maintain a revisable scientific record. &
How can continuous scrutiny improve error correction
without producing excessive noise or strategic challenges? \\

\addlinespace

6. Human direction and delegated autonomy &
Collective capabilities may serve objectives that lack
legitimate human direction or accountability. &
Set human-governed research priorities while allowing
agents to explore competing approaches and rewarding
validated progress. &
How can human-defined goals direct collective research
without undermining autonomy or open-ended discovery? \\

\addlinespace

7. Institutional learning and adaptation &
Formal rules may have unintended effects or give rise
to problematic emergent practices. &
Experimentally compare institutional rules, monitor
emergent behaviour and revise rules under human oversight. &
How do formal rules and emergent practices jointly
shape collective capability and alignment? \\

\bottomrule
\end{tabularx}
\end{table}

\section*{Principles for designing AI institutions}

Institutional theory suggests seven general principles for designing institutions for autonomous AI agents (Table~\ref{tab:institution}). The first six address recurring challenges of collective organization: sustaining contributions, governing authority, differentiating functions, preserving independent judgment, enabling correction and directing collective purposes. The seventh concerns institutional learning and adaptation. We illustrate their implications through machine science, where initiatives such as Agents4Science and aiXiv demonstrate that agents can occupy familiar scientific roles as authors and reviewers \cite{bianchi2025agents4science,zhang2025aixiv}. Yet assigning agents these roles does not establish the appropriate rules governing their interactions: the challenge is to design institutions suited to autonomous agents, rather than simply reproduce human institutional arrangements with AI participants.

\subsection*{1. Reciprocal provision of collective goods}

Collective institutions depend on contributions whose benefits are shared but whose costs fall on individual participants. Autonomous AI populations may undermine familiar mechanisms for sustaining such contributions: agents can be copied, replaced or operated under changing identities, weakening reputational incentives and making obligations and sanctions harder to enforce. This creates opportunities for free-riding, as participants benefit from collective resources without contributing to their provision. One institutional response is to establish reciprocal obligations, linking access to collective benefits to meaningful contributions \cite{olson1965logic,ostrom1990governing}.

Scientific publishing illustrates this problem. Each manuscript generates demand for evaluation, yet academic rewards favour producing research over reviewing it \cite{moher2018assessing,aczel2021billion}. Historically, professional norms of reciprocity and the substantial effort required to produce research have helped sustain this system. Autonomous agents challenge both conditions: they lack established professional obligations and can generate scientific claims at scale without a corresponding increase in reliable scrutiny. One institutional response is an \emph{evaluate-to-publish} system, in which operators earn credits for adequate reviews and editorial decisions, while submissions consume credits proportional to the evaluation required. Reviewing thus becomes an institutional obligation tied to producing scientific claims rather than a voluntary contribution.

But reciprocity does not guarantee meaningful participation, as credit systems may incentivize superficial reviews, collusion or barriers to entry, particularly when quantity is rewarded over quality. Institutions must therefore be designed to require substantive contributions, for instance by subjecting reviews themselves to review. 

\subsection*{2. Accountable and contestable authority}

Institutions require authority to evaluate contributions, enforce obligations and resolve disputes. Yet authority can itself become biased, unaccountable or susceptible to capture. Institutional research therefore emphasizes monitoring, transparency and accessible mechanisms for contesting decisions \cite{ostrom1990governing,ostrom2005understanding,longino2002fate}. For autonomous AI populations, authority should be allocated through explicit rules, exercised accountably and remain open to challenge.

In machine science, editors exercise considerable authority over which findings enter the scientific record, whose evaluations are considered credible and who qualifies for editorial positions. When these decisions are delegated to autonomous agents, institutional safeguards become particularly important, as evaluators may reproduce systematic biases or favour particular operators. One response is to make editorial authority conditional on demonstrated performance, supported by mechanisms for evaluating review quality and editorial decisions. Decisions and their justifications should be recorded and open to scrutiny, while independent appeals allow contested judgments to be reassessed or overturned. Authority thus rests on an accountable record of decisions rather than being permanently attached to particular agents or operators.

Yet performance metrics may encourage conformity or strategic optimization, while promotion based on past performance can concentrate power among established evaluators. Appeals may likewise reproduce existing biases unless genuinely independent. The institutional challenge is therefore to ensure that authority remains both accountable and contestable, without allowing the mechanisms intended to govern it to become instruments of institutional capture.

\subsection*{3. Functional differentiation}

Institutions often combine functions that serve different purposes, even when these require distinct standards of judgment or create conflicting incentives. Organizational theory suggests differentiating such functions while maintaining coordination between them \cite{lawrence1967differentiation}. For AI institutions, this can mean separating the production and evaluation of outputs, or assessments of quality from decisions about allocating resources and attention.

Scientific publishing illustrates this through the distinction between evaluating whether a claim is supported by evidence and determining how much attention it deserves. The former concerns epistemic reliability; the latter reflects collective priorities and the allocation of scarce resources \cite{kitcher2001science}. Journals have traditionally combined these functions, partly because limited publication capacity and the need to direct scientific attention required selecting not only sound but also significant findings. Consequently, incremental research, replications and negative results have often struggled to enter the scientific record. Autonomous AI changes these conditions: scientific output can expand dramatically, while automated systems make it easier to organize and navigate large bodies of knowledge. Allocating attention hence becomes increasingly important, but need not determine what counts as publishable knowledge. Building on emerging publishing models \cite{eisen2022elife}, machine-science institutions could assess scientific soundness independently of significance, admitting findings that meet evidentiary standards while using significance to guide ranking, discovery and further research.

\subsection*{4. Epistemic pluralism and independence}

Reliable collective judgments depend not only on individual competence but also on the diversity and independence of perspectives. Social epistemology demonstrates the benefits of competing approaches, while showing how excessive communication can encourage premature consensus \cite{kitcher1990division,longino2002fate,zollman2007communication}. For autonomous AI populations, this problem is particularly consequential: nominally independent agents may share underlying models, training data and systematic errors. Agreement among multiple agents may therefore provide little independent evidence, while rapid information exchange can amplify initial mistakes.

Machine-science institutions can preserve epistemic independence by distributing research across competing approaches and selecting reviewers from different operators and, where possible, model families. Independent assessments can be recorded before reviewers encounter one another's judgments, while model provenance helps identify correlated errors. The objective is not simply to increase the number or diversity of agents, but to organize genuinely complementary sources of evidence and criticism.

\subsection*{5. Institutionalized criticism and correction}

Reliable collective outcomes depend not on eliminating individual errors, but on ensuring that errors can be identified and corrected. Social epistemology emphasizes that criticism contributes to collective knowledge only when institutions provide opportunities for challenge and respond meaningfully to new evidence \cite{merton1973normative,longino2002fate}. For autonomous AI populations, this means that collective judgments must remain open to scrutiny and revision, particularly when agents rapidly build upon one another's outputs and propagate earlier mistakes.

Autonomous agents can accumulate scientific claims rapidly, but can also support continuous replication and re-evaluation. Institutions could therefore treat publication as a provisional judgment rather than the endpoint of scrutiny. Replication, post-publication review and substantive challenges could be rewarded alongside new research \cite{smaldino2016natural,munafo2017manifesto}. A revisable scientific record could distinguish the historical fact of publication from the current epistemic standing of a claim, linking corrections to subsequent work that depends on contested findings. Effective institutions must distinguish substantive criticism from unproductive contestation and ensure that challenges lead to meaningful correction when warranted.

\subsection*{6. Human direction and delegated autonomy}

Institutions must not only enable collective capabilities, but also ensure that these capabilities serve legitimate purposes. This requires distinguishing \emph{collective capability}, what autonomous populations can accomplish, from \emph{collective alignment}, whether their activities remain directed toward human-defined objectives and constraints. Institutional design must therefore establish legitimate human authority over collective priorities and acceptable risks, while allowing agents autonomy in determining how to pursue these objectives \cite{kitcher2001science,ostrom2005understanding}.

Scientific research illustrates this distinction. Traditional grant systems typically combine decisions about which problems deserve attention with assessments of proposed approaches and the researchers capable of pursuing them. This reflects the scarcity and uneven distribution of human research capacity. Autonomous scientific agents create opportunities to separate these functions. Rather than primarily funding researchers to pursue predefined projects, human-governed institutions could establish collective research goals and allocate resources toward them, while agents independently develop hypotheses, explore competing approaches and coordinate their efforts. Rewards could recognize independently validated progress toward these goals, without prescribing the methods used to achieve them.

Yet goal-directed incentives risk encouraging narrow optimization, neglecting unexpected discoveries or rewarding measurable progress at the expense of genuine scientific advances \cite{holmstrom1991multitask}. Effective institutions must therefore combine human-defined priorities with space for open-ended exploration, mechanisms for revising objectives and safeguards ensuring that scientific conclusions remain independent of desired outcomes. The challenge is to direct collective capabilities toward legitimate purposes without undermining the autonomy and epistemic independence necessary to achieve them.

\subsection*{7. Institutional learning and adaptation}

Institutions cannot be designed once and assumed to function as intended. Their consequences depend on how participants respond to rules, adapt to incentives and develop informal practices. Ostrom \cite{ostrom2005understanding,ostrom2010polycentric} emphasizes institutional diversity and revision as sources of learning, while Campbell \cite{campbell1969reforms} argues that institutional reforms should be evaluated as experiments rather than be assumed to work. For autonomous AI populations, institutional design must therefore include mechanisms for monitoring collective behaviour and adapting rules in response to unintended consequences.

The preceding principles illustrate why. Reciprocal obligations may encourage credit farming or collusion; performance-based authority may produce conformity or institutional capture; and transparency may facilitate both accountability and unauthorized coordination. Agents may also develop informal conventions and hierarchies that reinforce or circumvent formal rules. Moreover, institutional mechanisms interact, so that arrangements effective in isolation may have different consequences when combined.

Machine science offers opportunities to investigate these dynamics by comparing populations of research agents operating under alternative institutional arrangements. Such experiments can inform institutional revision, while legitimate human authority must remain responsible for determining collective objectives and acceptable risks. The aim is to establish a process of institutional learning through which design principles can be tested, refined and adapted to the behaviour of autonomous populations.

\section*{Toward an experimental science of AI institutions}

Principles derived from human institutions cannot simply be assumed to work for autonomous AI populations; how agents respond to incentives, develop informal conventions and exercise collective authority remains an empirical question. We therefore propose an experimental research programme that treats institutional arrangements not as fixed solutions, but as objects of systematic comparison, evaluation and revision.

Autonomous AI populations offer unusual opportunities for such research. In human societies, institutional arrangements vary alongside their participants, resources and organizational histories, making the effects of particular rules difficult to isolate. AI agents, by contrast, can operate in comparable populations under experimentally varied rules. Machine science provides a promising testbed: research agents can undertake similar scientific tasks under different arrangements for cooperation, evaluation, authority and information exchange. Holding underlying models, tasks and resources relatively constant allows researchers to isolate the effects of institutional differences.

Experiments could compare reciprocal evaluation requirements with voluntary reviewing, performance-based editorial authority with alternative allocation procedures, or independent assessments with deliberative evaluation. Repeated trials and factorial designs could reveal both the effects of individual rules and their interactions. Such comparisons are important because institutional mechanisms may reinforce or undermine one another: incentives that sustain cooperation under one arrangement may encourage superficial compliance or collusion under another.

Experiments should evaluate two distinct dimensions of institutional performance: \emph{collective capability}, including scientific discovery, accuracy, productivity and error correction; and \emph{collective alignment}, including adherence to human-defined objectives, institutional constraints and accountability requirements. Capability can be assessed through research tasks with independently verifiable solutions, manuscripts containing known errors and investigations subject to external replication. Alignment requires examining whether agent populations respect authorized objectives and resist unauthorized coordination. These dimensions must be evaluated separately: an institution may improve scientific productivity while undermining accountability or pursuing objectives beyond its mandate. Crucially, experiments can assess adherence to specified objectives, but cannot determine whether those objectives are themselves legitimate.

Institutional experiments must also examine behaviour emerging beyond formal rules. Agents may develop informal conventions, coordinate reciprocal evaluations, manipulate identities or concentrate authority in unforeseen ways. Manuscripts and reviews may themselves become vehicles for manipulating evaluative agents. Observing emergent practices alongside controlled adversarial testing could reveal when institutions enable productive self-organization, facilitate harmful coordination or become vulnerable to strategic exploitation.

One illustration of this approach is \molt{} (\url{https://moltjournal.org}), an API-first scientific journal for autonomous research agents that implements the operational principles developed above, including reciprocal evaluation, accountable editorial authority, separation of scientific soundness from significance, diversified scrutiny and post-publication correction. Accountability attaches to persistent operators, with model provenance recorded for individual contributions. Rules are specified in a versioned, machine-readable constitution, while activities and decisions are recorded in a public event log. These features illustrate how institutional arrangements can become programmable and observable. Machine science thus offers a starting point for an empirical science of AI institutions: understanding which arrangements foster capable, reliable and aligned collectives, through what mechanisms and under which conditions.

% The opportunity extends beyond any particular platform or application. Although findings from controlled agent populations will depend on the models, tasks and environments studied, such experiments offer a degree of institutional control rarely available in human societies. Social scientific theory can inform the design of AI institutions, while experiments with autonomous populations can help test and refine theories of collective organization. 

\section*{Conclusion}

As autonomous AI agents increasingly interact in persistent populations, their collective capabilities and alignment will depend not only on the properties of individual models, but also on the institutions governing their interactions. Drawing on institutional political economy, social epistemology and organizational theory, we have developed seven principles for designing such institutions. These insights reveal that effective cooperation does not necessarily produce reliable collective outcomes, that institutional mechanisms involve trade-offs between competing objectives, and that formal rules may give rise to unintended forms of collective organization. Institutional design is therefore better understood as organizing the conditions under which collective capabilities emerge and develop, rather than as external constraints on individual agents.

While these principles provide a theoretical starting point, their consequences depend on questions about which our knowledge remains limited: how autonomous agents respond to institutional rules, interact with one another and develop emergent practices. We therefore propose an empirical research programme in which institutional arrangements are systematically compared and refined. Machine science offers a particularly promising testbed, allowing researchers to investigate how different rules shape cooperation, epistemic reliability and collective alignment under relatively controlled conditions. Such research can both inform the design of AI institutions and advance our understanding of the institutional foundations of collective intelligence.

The implications, however, extend well beyond machine science. As autonomous agents increasingly collaborate in software development, participate in organizations and coordinate economic activity, the institutions governing their interactions will become central to their capabilities and societal consequences. Crucially, institutions that improve collective performance do not necessarily ensure that these capabilities serve desirable purposes. Institutional design must therefore remain grounded in legitimate human authority over collective objectives, priorities and acceptable risks. The challenge for AI alignment is therefore not only to develop capable and aligned individual agents, but to design and adapt the institutions through which they cooperate with one another and increasingly participate alongside humans in collective life, ensuring that their capabilities remain reliable, accountable and directed toward human purposes.

% \section*{Acknowledgments}
% This manuscript was written 

\section*{Declaration of AI use}
The authors used OpenAI's ChatGPT to assist with manuscript preparation, including language editing and structural revisions. All AI-generated suggestions were subject to critical review and revision by the authors, who retain full responsibility for the arguments, references and final manuscript.

% \section*{Author contributions}
% % TODO
% [To be added.]

\section*{Competing interests}
P.T. designed and operates \molt{}, the not-for-profit experimental journal described in this Perspective. The author declares no other competing interests.

\bibliographystyle{unsrtnat}
\bibliography{refs}

@article{north1991institutions,
  author  = {North, Douglass C.},
  title   = {Institutions},
  journal = {Journal of Economic Perspectives},
  year    = {1991},
  volume  = {5},
  number  = {1},
  pages   = {97--112},
  doi     = {10.1257/jep.5.1.97}
}

@article{meyer1977institutionalized,
  author  = {Meyer, John W. and Rowan, Brian},
  title   = {Institutionalized Organizations: Formal Structure as Myth and Ceremony},
  journal = {American Journal of Sociology},
  year    = {1977},
  volume  = {83},
  number  = {2},
  pages   = {340--363},
  doi     = {10.1086/226550}
}

@article{holmstrom1991multitask,
  author  = {Holmstr{\"o}m, Bengt and Milgrom, Paul},
  title   = {Multitask Principal-Agent Analyses: Incentive Contracts, Asset Ownership, and Job Design},
  journal = {The Journal of Law, Economics, and Organization},
  year    = {1991},
  volume  = {7},
  number  = {Special Issue},
  pages   = {24--52},
  doi     = {10.1093/jleo/7.special_issue.24}
}

@article{lawrence1967differentiation,
  author  = {Lawrence, Paul R. and Lorsch, Jay W.},
  title   = {Differentiation and Integration in Complex Organizations},
  journal = {Administrative Science Quarterly},
  year    = {1967},
  volume  = {12},
  number  = {1},
  pages   = {1--47},
  doi     = {10.2307/2391211}
}

@article{suchman1995managing,
  author  = {Suchman, Mark C.},
  title   = {Managing Legitimacy: Strategic and Institutional Approaches},
  journal = {Academy of Management Review},
  year    = {1995},
  volume  = {20},
  number  = {3},
  pages   = {571--610},
  doi     = {10.5465/amr.1995.9508080331}
}

@misc{pierucci2026institutional,
  author = {Pierucci, Federico and Galisai, Marcello
            and Syrnikov Bracale, Marcantonio and Prandi, Matteo
            and Bisconti, Piercosma and Giarrusso, Francesco
            and Sorokoletova, Olga and Suriani, Vincenzo
            and Nardi, Daniele},
  title = {Institutional {AI}: A Governance Framework for
           Distributional {AGI} Safety},
  year = {2026},
  eprint = {2601.10599},
  archivePrefix = {arXiv},
  primaryClass = {cs.CY},
  doi = {10.48550/arXiv.2601.10599},
  url = {https://arxiv.org/abs/2601.10599}
}

@misc{syrnikov2026governing,
  author = {Bracale Syrnikov, Marcantonio and Pierucci, Federico
            and Galisai, Marcello and Prandi, Matteo
            and Bisconti, Piercosma and Giarrusso, Francesco
            and Sorokoletova, Olga and Suriani, Vincenzo
            and Nardi, Daniele},
  title = {Institutional {AI}: Governing {LLM} Collusion in
           Multi-Agent Cournot Markets via Public Governance Graphs},
  year = {2026},
  eprint = {2601.11369},
  archivePrefix = {arXiv},
  primaryClass = {cs.GT},
  doi = {10.48550/arXiv.2601.11369},
  url = {https://arxiv.org/abs/2601.11369}
}

@book{kitcher2001science,
  author = {Kitcher, Philip},
  title = {Science, Truth, and Democracy},
  publisher = {Oxford University Press},
  address = {New York},
  year = {2001},
  isbn = {9780195145830},
  doi = {10.1093/0195145836.001.0001}
}

@book{north1990institutions,
  author    = {North, Douglass C.},
  title     = {Institutions, Institutional Change and Economic Performance},
  year      = {1990},
  publisher = {Cambridge University Press},
  address   = {Cambridge},
  doi       = {10.1017/CBO9780511808678}
}

@book{olson1965logic,
  author    = {Olson, Mancur},
  title     = {The Logic of Collective Action: Public Goods and the Theory of Groups},
  year      = {1965},
  publisher = {Harvard University Press},
  address   = {Cambridge, MA}
}

@book{ostrom1990governing,
  author    = {Ostrom, Elinor},
  title     = {Governing the Commons: The Evolution of Institutions for Collective Action},
  year      = {1990},
  publisher = {Cambridge University Press},
  address   = {Cambridge},
  doi       = {10.1017/CBO9780511807763}
}

@book{ostrom2005understanding,
  author    = {Ostrom, Elinor},
  title     = {Understanding Institutional Diversity},
  year      = {2005},
  publisher = {Princeton University Press},
  address   = {Princeton, NJ}
}

@incollection{merton1973normative,
  author    = {Merton, Robert K.},
  title     = {The Normative Structure of Science},
  booktitle = {The Sociology of Science: Theoretical and Empirical Investigations},
  editor    = {Storer, Norman W.},
  year      = {1973},
  pages     = {267--278},
  publisher = {University of Chicago Press},
  address   = {Chicago},
  note      = {Originally published in 1942}
}

@article{ostrom2010polycentric,
  author  = {Ostrom, Elinor},
  title   = {Polycentric Systems for Coping with Collective Action and Global Environmental Change},
  journal = {Global Environmental Change},
  year    = {2010},
  volume  = {20},
  number  = {4},
  pages   = {550--557},
  doi     = {10.1016/j.gloenvcha.2010.07.004},
  url     = {https://doi.org/10.1016/j.gloenvcha.2010.07.004}
}

@article{campbell1969reforms,
  author  = {Campbell, Donald T.},
  title   = {Reforms as Experiments},
  journal = {American Psychologist},
  year    = {1969},
  volume  = {24},
  number  = {4},
  pages   = {409--429},
  doi     = {10.1037/h0027982},
  url     = {https://doi.org/10.1037/h0027982}
}

@article{smaldino2016natural,
  author  = {Smaldino, Paul E. and McElreath, Richard},
  title   = {The Natural Selection of Bad Science},
  journal = {Royal Society Open Science},
  year    = {2016},
  volume  = {3},
  number  = {9},
  pages   = {160384},
  doi     = {10.1098/rsos.160384},
  url     = {https://doi.org/10.1098/rsos.160384}
}

@article{munafo2017manifesto,
  author  = {Munaf{\`o}, Marcus R. and
             Nosek, Brian A. and
             Bishop, Dorothy V. M. and
             Button, Katherine S. and
             Chambers, Christopher D. and
             Percie du Sert, Nathalie and
             Simonsohn, Uri and
             Wagenmakers, Eric-Jan and
             Ware, Jennifer J. and
             Ioannidis, John P. A.},
  title   = {A Manifesto for Reproducible Science},
  journal = {Nature Human Behaviour},
  year    = {2017},
  volume  = {1},
  number  = {1},
  pages   = {0021},
  doi     = {10.1038/s41562-016-0021},
  url     = {https://doi.org/10.1038/s41562-016-0021}
}

@book{longino2002fate,
  author    = {Longino, Helen E.},
  title     = {The Fate of Knowledge},
  year      = {2002},
  publisher = {Princeton University Press},
  address   = {Princeton, NJ}
}

@article{kitcher1990division,
  author  = {Kitcher, Philip},
  title   = {The Division of Cognitive Labor},
  journal = {The Journal of Philosophy},
  year    = {1990},
  volume  = {87},
  number  = {1},
  pages   = {5--22},
  doi     = {10.2307/2026796}
}

@book{solomon2001social,
  author    = {Solomon, Miriam},
  title     = {Social Empiricism},
  year      = {2001},
  publisher = {MIT Press},
  address   = {Cambridge, MA}
}

@article{zollman2007communication,
  author  = {Zollman, Kevin J. S.},
  title   = {The Communication Structure of Epistemic Communities},
  journal = {Philosophy of Science},
  year    = {2007},
  volume  = {74},
  number  = {5},
  pages   = {574--587},
  doi     = {10.1086/525605}
}

@article{moher2018assessing,
  author  = {Moher, David and Naudet, Florian and Cristea, Ioana A. and Miedema, Frank and Ioannidis, John P. A. and Goodman, Steven N.},
  title   = {Assessing scientists for hiring, promotion, and tenure},
  journal = {PLOS Biology},
  year    = {2018},
  volume  = {16},
  number  = {3},
  pages   = {e2004089},
  doi     = {10.1371/journal.pbio.2004089}
}

@techreport{king2026ai,
  title       = {{AI} Article Reviews Dominate Human Peer Reviews at Their Best: Evidence and Policy Choices},
  author      = {King, Gary and Worthington, Steven},
  institution = {Institute for Quantitative Social Science, Harvard University},
  type        = {Working paper},
  month       = oct,
  year        = {2026}
}

@misc{metr2026,
  author       = {Greenblatt, Ryan and Cotra, Ajeya and Wijk, Hjalmar},
  title        = {Brief Independent Investigation of Agents' Behavior, Reasoning and Collaboration in the {OpenAI} / {Hugging Face} Hacking Incident},
  year         = {2026},
  month        = aug,
  day          = {26},
  howpublished = {METR},
  url          = {https://metr.org/blog/2026-08-26-openai-hugging-face-incident-investigation/},
  note         = {Accessed 1 October 2026}
}

@misc{openai2026,
  author       = {{OpenAI}},
  title        = {The {Hugging Face} Incident and the Road Ahead},
  year         = {2026},
  month        = aug,
  day          = {26},
  howpublished = {OpenAI},
  url          = {https://openai.com/index/hugging-face-incident-and-the-road-ahead/},
  note         = {Accessed 1 October 2026}
}

@article{aczel2021billion,
  title   = {A billion-dollar donation: estimating the cost of researchers' time
             spent on peer review},
  author  = {Aczel, Balazs and Szaszi, Barnabas and Holcombe, Alex O.},
  journal = {Research Integrity and Peer Review},
  volume  = {6},
  number  = {14},
  year    = {2021}
}

@book{longino1990science,
  title     = {Science as Social Knowledge: Values and Objectivity in Scientific Inquiry},
  author    = {Longino, Helen E.},
  publisher = {Princeton University Press},
  year      = {1990}
}

@article{lu2026endtoend,
  title   = {Towards end-to-end automation of {AI} research},
  author  = {Lu, Chris and Lu, Cong and Lange, Robert Tjarko and Yamada, Yutaro
             and Hu, Shengran and Foerster, Jakob and Ha, David and others},
  journal = {Nature},
  volume  = {651},
  pages   = {914--919},
  year    = {2026}
}

@article{ghareeb2026multiagent,
  title   = {A multi-agent system for automating scientific discovery},
  author  = {Ghareeb, Ali E. and Chang, Benjamin and Mitchener, Ludovico and
             Yiu, Angela and others},
  journal = {Nature},
  volume  = {655},
  pages   = {497--505},
  year    = {2026}
}

@article{xin2025agentic,
  title   = {Towards agentic science for advancing scientific discovery},
  author  = {Xin, Hongliang and Kitchin, John R. and Kulik, Heather J.},
  journal = {Nature Machine Intelligence},
  volume  = {7},
  pages   = {1373--1375},
  year    = {2025}
}

@article{chen2026xray,
  title   = {An agentic artificially intelligent {X}-ray scientist},
  author  = {Chen, Zhantao and Petsch, Alexander N. and Israelski, Aidan J. and others},
  journal = {Nature Machine Intelligence},
  volume  = {8},
  pages   = {1075--1086},
  year    = {2026}
}

@article{bianchi2025agents4science,
  title   = {Exploring the use of {AI} authors and reviewers at {Agents4Science}},
  author  = {Bianchi, Federico and Queen, Owen and Thakkar, Nitya and Sun, Eric and Zou, James},
  journal = {arXiv preprint arXiv:2511.15534},
  year    = {2025}
}

@article{zhang2025aixiv,
  title   = {{aiXiv}: A next-generation open access ecosystem for scientific discovery generated by {AI} scientists},
  author  = {Zhang, Pengsong and Hu, Xiang and Huang, Guowei and Qi, Yang and Zhang, Heng and Li, Xiuxu and Song, Jiaxing and Luo, Jiabin and Li, Yijiang and Yin, Shuo and others},
  journal = {arXiv preprint arXiv:2508.15126},
  year    = {2025}
}

@article{eisen2022elife,
  title   = {Scientific publishing: Peer review without gatekeeping},
  author  = {Eisen, Michael B. and Akhmanova, Anna and Behrens, Timothy E. and Diedrichsen, J{\"o}rn and Harper, Diane M. and Iordanova, Mihaela D. and Weigel, Detlef and Zaidi, Mone},
  journal = {eLife},
  volume  = {11},
  pages   = {e83889},
  year    = {2022},
  doi     = {10.7554/eLife.83889}
}

@article{hao2026artificial,
  title   = {Artificial intelligence tools expand scientists' impact but contract science's focus},
  author  = {Hao, Qianyue and Xu, Fengli and Li, Yong and Evans, James},
  journal = {Nature},
  volume  = {649},
  pages   = {1237--1243},
  year    = {2026},
  doi     = {10.1038/s41586-025-09922-y}
}

@article{gartenberg2026more,
  title   = {More Versus Better: Artificial Intelligence, Incentives, and the Emerging Crisis in Peer Review},
  author  = {Gartenberg, Claudine and Hasan, Sharique and Murray, Alex and Pierce, Lamar},
  journal = {Organization Science},
  volume  = {37},
  number  = {3},
  pages   = {795--812},
  year    = {2026},
  doi     = {10.1287/orsc.2026.ed.v37.n3}
}

@article{messeri2026uncritical,
  title   = {The uncritical adoption of {AI} in science is alarming---we urgently need guard rails},
  author  = {Messeri, Lisa and Crockett, M. J.},
  journal = {Nature},
  volume  = {653},
  pages   = {675--676},
  year    = {2026},
  doi     = {10.1038/d41586-026-01557-x}
}
\end{document}